\documentclass[11pt,letterpaper,draftcls,oneside,peerreview,onecolumn]{IEEEtran}
\usepackage{amsfonts}
\usepackage{amssymb}
\usepackage{amsmath}
\usepackage[dvips]{graphicx}
\usepackage{cite}

\begin{document}

\title{Adaptive Subcarrier PSK Intensity Modulation in Free Space Optical
Systems}
\author{Nestor D. Chatzidiamantis,~\IEEEmembership{Student~Member,~IEEE,}
Athanasios S. Lioumpas, \and ~\IEEEmembership{Student~Member,~IEEE,} George
K. Karagiannidis,~\IEEEmembership{Senior~Member,~IEEE,}\newline~and~Shlomi
Arnon,~\IEEEmembership{Senior~Member,~IEEE}\thanks{%
N. D. Chatzidiamantis, A. S. Lioumpas and G. K. Karagiannidis are with the
Wireless Communications Systems Group (WCSG), Department of Electrical and
Computer Engineering, Aristotle University of Thessaloniki, GR-54124
Thessaloniki, Greece (e-mails: \{nestoras, alioumpa, geokarag\}@auth.gr).}
\thanks{%
S. Arnon is with the Satellite and Wireless Communication Laboratory,
Department of Electrical and Computer Engineering, Ben-Gurion University of
the Negev, Beer-Sheva IL-84105, Israel (e-mail: shlomi@ee.bgu.ac.il).}}
\pubid{}
\specialpapernotice{}
\maketitle

\begin{abstract}
We propose an adaptive transmission technique for free space optical (FSO)
systems, operating in atmospheric turbulence and employing subcarrier phase
shift keying (S-PSK) intensity modulation. Exploiting the constant envelope
characteristics of S-PSK, the proposed technique offers efficient
utilization of the FSO channel capacity by adapting the modulation order of
S-PSK, according to the instantaneous state of turbulence induced fading and
a pre-defined bit error rate (BER) requirement. Novel expressions for the
spectral efficiency and average BER of the proposed adaptive FSO system are
presented and performance investigations under various turbulence conditions
and target BER requirements are carried out. Numerical results indicate that
significant spectral efficiency gains are offered without increasing the
transmitted average optical power or sacrificing BER requirements, in
moderate-to-strong turbulence conditions. Furthermore, the proposed variable
rate transmission technique is applied to multiple input multiple output
(MIMO) FSO systems, providing additional improvement in the achieved
spectral efficiency as the number of the transmit and/or receive apertures
increases.
\end{abstract}


\markboth{Submitted to \textit{IEEE Transactions on Communications}}
{Murray and Balemi: Using the Document Class IEEEtran.cls}%
\setcounter{page}{0}\newpage

\section{Introduction}

Free Space Optical (FSO) communication is a wireless technology, which has
recently attracted considerable interest within the research community,
since it can be advantageous for a variety of applications \cite{B:Andrews}{-%
}\nocite{J:Arn1}\cite{J:Chan_inv}. However, despite its significant
advantages, the widespread deployment of FSO systems is limited by their
high vulnerability to adverse atmospheric conditions \cite{B:Karp}. Even in
a clear sky, due to inhomogeneities in temperature and pressure changes, the
refractive index of the atmosphere varies stochastically and results in
atmospheric turbulence. This causes rapid fluctuations at the intensity of
the received optical signal, known as turbulence-induced fading, that
severely affect the reliability and/or communication rate provided by the
FSO link.

Over the last years, several fading mitigation techniques have been proposed
for deployment in FSO links to combat the degrading effects of atmospheric
turbulence. Error control coding (ECC) in conjunction with interleaving has
been investigated in \cite{J:Kah4} and \cite{J:Uysal1}. Although this
technique is known in the Radio Frequency (RF) literature to provide an
effective time-diversity solution to rapidly-varying fading channels, its
practical use in FSO links is limited due to the large-size interleavers%
\footnote{%
For the signalling rates of interest (typically of the the order of Gbps),
FSO channels exhibit slow fading, since the correlation time of turbulence
is of the order of 10$^{-3}$ to 10$^{-2}$ seconds.} required to achieve the
promising coding gains theoretically available. Maximum Likelihood Sequence
Detection (MLSD), which has been proposed in \cite{J:Kah1}, efficiently
exploits the channel's temporal characteristics; however it suffers from
extreme computational complexity and therefore in practice, only suboptimal
MLSD solutions can be employed \cite{J:SchoberTCOM}\nocite{C:Sch_2}-\cite%
{C:Sch_4}. Of particular interest is the application of spatial diversity to
FSO systems, i.e., transmission and/or reception through multiple apertures,
since significant performance gains are offered by taking advantage the
additional degrees of freedom in the spatial dimension \cite{J:Navid}-\cite%
{J:Tsif}. Nevertheless, increasing the number of apertures, increases the
cost and the overall physical size of the FSO systems.

Another promising solution is the employment of adaptive transmission, a
well known technique employed in RF systems \cite{J:Aluini_A}. By varying
basic transmission parameters according to the channel's fading intensity,
adaptive transmission takes advantage of the time varying nature of
turbulence, allowing higher data rates to be transmitted under favorable
turbulence conditions. Thus, the spectral efficiency of the FSO link can be
improved without wasting additional optical power or sacrificing performance
requirements. The concept of adaptive transmission was first introduced in
the context of FSO systems in \cite{J:Levitt}, where an adaptive scheme that
varied the period of the transmitted binary Pulse Position modulated (BPPM)
symbol was studied. Since then, various adaptive FSO systems have been
proposed. In \cite{C:Uysal_Glob03}, a variable rate FSO system employing
adaptive Turbo-based coding schemes in conjunction with On-Off keying
modulation was investigated, while in \cite{J:Arn_3} an adaptive power
scheme was suggested for reducing the average power consumption in constant
rate optical satellite-to-earth links. Recently, in \cite{J:Djordj}, an
adaptive transmission scheme that varied both the power and the modulation
order of a FSO system with pulse amplitude modulation (PAM), has been
studied.

In this work, we propose an adaptive modulation scheme for FSO systems
operating in turbulence, using an alternative type of modulation; subcarrier
phase shift keying intensity modulation (S-PSK). S-PSK refers to the
transmission of PSK modulated RF signals, after being properly biased%
\footnote{%
Since optical intensity must satisfy the non-negativity constraint, a proper
DC bias must be added to the RF electrical signal in order to prevent
clipping and distortion in the optical domain.}, through intensity
modulation direct detection (IM/DD) optical systems and its employment can
be advantageous, since:

\begin{itemize}
\item in the presence of turbulence, it offers increased demodulation
performance compared to PAM signalling \cite{C:Liu}\nocite{C:Liu_1}-\cite%
{J:Liu},

\item due to its constant envelope characteristic, both the average and peak
optical power are constant in every symbol transmitted, and,

\item it allows RF signals to be directly transmitted through FSO links
providing protocol transparency in heterogeneous wireless networks \cite%
{J:Arnon_Wimax}-\nocite{C:Cvijetic_1}\cite{J:Arnon_cell}.
\end{itemize}

Taking into consideration that the bias signal required by S-PSK in order to
satisfy the non-negativity requirement is independent of the modulation
order, we present a novel variable rate transmission strategy that is
implemented through the modification of the modulation order of S-PSK,
according to the instantaneous turbulence induced fading and a pre-defined
value of Bit Error Rate (BER). The performance of the proposed variable rate
transmission scheme is investigated, in terms of spectral efficiency and
BER, under different degrees of turbulence strength, and is further compared
to non-adaptive modulation and the upper capacity bound provided by \cite%
{J:Kah6}. Moreover, an application to multiple input multiple output (MIMO)
FSO systems employing equal gain combining (EGC) at the receiver is provided
and the performance of the presented transmission policy is evaluated for
various MIMO deployments.

The remainder of the paper is organized as follows. In Section \ref{Non
Adaptive}, the non-adaptive S-PSK FSO system model is described and its
performance in the presence of turbulence induced fading is investigated. In
section \ref{Adaptive Modulation}, the adaptive S-PSK strategy is presented,
deriving expressions for its spectral efficiency and BER performance, and an
application to MIMO FSO systems is further provided. Section \ref{Results}
discusses some numerical results and useful concluding remarks are drawn in
section \ref{Conclusions}.

\emph{Notations}: $\mathrm{E}\left\{ \cdot \right\} $ denotes statistical
expectation; $N\left( \mu ,\sigma ^{2}\right) $ denotes Gaussian
distribution with mean $\mu $ and variance $\sigma ^{2}$.

\section{Non-Adaptive Subcarrier PSK Intensity Modulation\label{Non Adaptive}%
}

We consider an IM/DD FSO system which uses a subcarrier signal for the
modulation of the optical carrier's intensity and operates over the
atmospheric turbulence induced fading channel.

\subsection{System and Channel Model}

On the transmitter end, we assume that the RF subcarrier signal is modulated
by the data sequence using PSK. Moreover a proper DC bias is added in order
to ensure that the transmitted waveform always satisfies the non-negativity
input constraint. Hence, the transmitted optical power can be expressed as
\begin{equation}
P_{t}\left( t\right) =P\left[ 1+\mu s\left( t\right) \right]
\label{transmited_power}
\end{equation}%
where $P$ is the average transmitted optical power and $\mu $ is the
modulation index $\left( 0<\mu <1\right) $ which ensures that the laser
operates in its linear region and avoids over-modulation induced clipping.
Further, $s\left( t\right) $ is the output of the electrical PSK modulator
which can be written as
\begin{equation}
s\left( t\right) =\sum_{k}g\left( t-kT\right) \cos \left( 2\pi f_{c}t+\phi
_{k}\right)  \label{PSK_symbol}
\end{equation}%
where $f_{c}$ is the frequency of the RF subcarrier signal, $T$ is the
symbol's period, $g\left( t\right) $ is the shaping pulse, $\phi _{k}\in %
\left[ 0,...,\left( M-1\right) \frac{\pi }{M}\right] $ is the phase of the $%
k $th transmitted symbol and $M$ is the modulation order.

On the receiver's end, the optical power which is incident on the
photodetector is converted into an electrical signal through direct
detection. We assume operation in the high signal-to-noise ratio (SNR)
regime where the shot noise caused by ambient light is dominant and
therefore Gaussian noise model is used as a good approximation of the
Poisson photon counting detection model \cite{J:Kah1}.

After removing the DC bias and demodulating through a standard RF PSK
demodulator, the electrical signal sampled during the $k$th symbol interval,
which is obtained at the output of the receiver, is expressed as
\begin{equation}
r\left[ k\right] =\mu \eta \sqrt{\frac{E_{g}}{2}}PI\left[ k\right] s\left[ k%
\right] +n\left[ k\right]  \label{baseband_received}
\end{equation}%
where $\eta $\ corresponds to the receiver's optical-to-electrical
efficiency, $s\left[ k\right] =\cos \phi _{k}-j\sin \phi _{k}$, $E_{g}$ is
the energy of the shaping pulse and $n\left[ k\right] $ is the zero mean
circularly symmetric complex Gaussian noise component with $\mathrm{E}%
\left\{ n\left[ k\right] n^{\ast }\left[ k\right] \right\} =2\sigma
_{n}^{2}=N_{o}$. Furthermore, $I\left[ k\right] $ represents
turbulence-induced fading coefficient during the $k$th symbol interval and
is given by%
\begin{equation}
I\left[ k\right] =I_{o}\exp \left( 2x\left[ k\right] \right)  \label{Eq:2}
\end{equation}%
where $I_{o}$ denotes the signal light intensity without turbulence and $x%
\left[ k\right] $ is a normally distributed random variable with mean $m_{x}$
and variance $\sigma _{x}^{2}$, i.e. $f_{x\left[ k\right] }\left( x\right)
=N\left( m_{x},\sigma _{x}^{2}\right) $. Hence $I\left[ k\right] $ follows a
lognormal distribution with probability density function (PDF) provided by%
\begin{equation}
f_{I\left[ k\right] }\left( I\right) =\frac{1}{2I}\frac{1}{\sqrt{2\pi \sigma
_{x}^{2}}}\exp \left( -\frac{\left( \ln \left( \frac{I}{I_{o}}\right)
-2m_{x}\right) ^{2}}{8\sigma _{x}^{2}}\right)  \label{Eq:3}
\end{equation}%
To ensure that the fading does not attenuate or amplify the average power,
we normalize the fading coefficients such that $\mathrm{E}\left\{ \left\vert
\frac{I\left[ k\right] }{I_{o}}\right\vert \right\} =1$. Doing so requires
the choice of $m_{x}=-\sigma _{x}^{2}$ \cite{J:Navid}. Moreover, without
loss of generality, it is assumed that $I_{o}=1$.

Atmospheric turbulence results in a very slowly-varying fading in FSO
systems. For the signalling rates of interest ranging from hundreds to
thousands of Mbps \cite{J:Heatley}, the fading coefficient can be considered
as constant over hundred of thousand or millions of consecutive symbols,
since the coherence time of the channel is about 1-100ms \cite{J:Lee}.
Hence, it is assumed that turbulence induced fading remains constant over a
block of $K$ symbols (block fading channel), and therefore we drop the time
index $k$, i.e.%
\begin{equation}
I=I\left[ k\right] ,~~k=1,...K
\end{equation}%
It should be noted that in the analysis that follows, it is further assumed
that the information message is long enough to reveal the long-term ergodic
properties of the turbulence process.

The instantaneous electrical SNR is defined as%
\begin{equation}
\gamma =\frac{\mu ^{2}\eta ^{2}P^{2}E_{s}I^{2}}{N_{o}}
\end{equation}%
while the average electrical SNR is given by%
\begin{equation}
\bar{\gamma}=\frac{\mu ^{2}\eta ^{2}P^{2}E_{s}}{N_{o}}  \label{average_SNR}
\end{equation}%
with $E_{s}=\frac{E_{g}}{2}$.

\subsection{BER Performance}

The BER performance of the FSO system under consideration depends on the
statistics of atmospheric turbulence and the modulation order.

When $M=2$ (BPSK), the conditioned on the fading coefficient, $I$, BER is
given by \cite{J:Liu}%
\begin{equation}
P_{b}\left( 2,I\right) =Q\left( I\sqrt{2\bar{\gamma}}\right)  \label{BER_M=2}
\end{equation}%
while for $M>2$ the following approximation can be used \cite{B:Proakis}%
\begin{equation}
P_{b}\left( M,I\right) \approx \frac{2}{\log _{2}M}Q\left( I\sqrt{2\bar{%
\gamma}}\sin \frac{\pi }{M}\right)  \label{BER_M>2}
\end{equation}%
where $Q\left( \cdot \right) $ is the Gaussian Q-function defined as $%
Q\left( x\right) =\frac{1}{\sqrt{2\pi }}\int_{x}^{\infty }e^{-\frac{t^{2}}{2}%
}dx$. Hence, the average BER will be obtained by averaging over the
turbulence PDF, i.e.%
\begin{equation}
\bar{P}_{b}\left( M\right) =\int_{0}^{\infty }P_{b}\left( M,I\right)
f_{I}\left( I\right) dI  \label{uncond_BER}
\end{equation}%
which can be efficiently evaluated using the Gauss-Hermitte quadrature
formula \cite{J:Navid}.

\subsection{Capacity Upper Bound}

Using the trigonometric moment space method, an upper bound for the capacity
of optical intensity channels when multiple subcarrier modulation is
employed, has been derived in \cite{J:Kah6}. By applying these results to
the FSO system under consideration (one subcarrier), the conditioned on the
fading coefficient, $I$, capacity can be upper bounded by
\begin{equation}
C_{up}\left( I\right) =\frac{W}{2}\left[ \log _{2}\pi +\log _{2}\left( \frac{%
\mu ^{2}\eta ^{2}P^{2}E_{g}I^{2}}{\pi eN_{o}}\right) +o\left( \sigma
_{n}\right) \right]  \label{capacity_upper_bound}
\end{equation}%
where $W$ denotes the electrical bandwidth and $o\left( \sigma _{n}\right) $
represents the capacity residue which vanishes exponentially as $\sigma
_{n}\rightarrow 0$. Hence at high values of electrical SNR, (\ref%
{capacity_upper_bound}) can be approximated by%
\begin{equation}
C_{up}\left( I\right) \approx \frac{W}{2}\log _{2}\left( \frac{\bar{\gamma}%
I^{2}}{e}\right) .  \label{capacity_upp_appr}
\end{equation}

The unconditional approximative capacity upper bound will be obtained by
averaging (\ref{capacity_upp_appr}) over the fading distribution, i.e.%
\begin{equation}
C_{up}\approx \frac{W}{2}\int_{0}^{\infty }\log _{2}\left( \frac{\bar{\gamma}%
I^{2}}{e}\right) f_{I}\left( I\right) dI,  \label{uncond_cap}
\end{equation}%
which, according to the Appendix, can be written in closed form as
\begin{equation}
C_{up}\approx \frac{W}{2}\left( \log _{2}\left( \frac{\bar{\gamma}}{e}%
\right) -\frac{4\sigma _{x}^{2}}{\ln 2}\right)  \label{uncond_cap2}
\end{equation}%
and will be used as a benchmark in the analysis that follows.

\section{Adaptive Modulation Strategy\label{Adaptive Modulation}}

In this section we introduce an adaptive transmission strategy that improves
the spectral efficiency of S-PSK FSO systems, without increasing the
transmitted average optical power or sacrificing the performance
requirements.

\subsection{Mode of Operation}

By inserting pilot symbols at the beginning of a block of symbols\footnote{%
Taking into consideration the length of a transmitting block of symbols, the
insertion of pilot symbols will not cause significant overhead.}, the
receiver accurately estimates the instantaneous channel's fading state, $I$,
which is experienced by the remaining symbols of the block. Based on this
estimation, a decision device at the receiver selects the modulation order
to be used for transmitting the non-pilot symbols of the block, configures
the electrical demodulator accordingly and informs the adaptive PSK
transmitter about that decision via a reliable RF feed back path.

The objective of the above described transmission technique is to maximize
the number of bits transmitted per symbol interval, by using the largest
possible modulation order under the target BER requirement $P_{o}$. Hence
the problem is formulated as
\begin{equation}
\begin{split}
\max_{M}~\log _{2}M& \\
\text{s.t.}~P_{b}\left( M,I\right) & \leq P_{o}
\end{split}
\label{optimization problem}
\end{equation}%
In practice, the modulation order will be selected from $N$ available ones,
i.e. $\left\{ M_{1},M_{2},...,M_{N}\right\} $, depending on the values of $I$
and $P_{o}$. Specifically, the range of the values of the fading term is
divided in $\left( N+1\right) $ regions and each region is associated with
the modulation order, $M_{j}$, according to the rule%
\begin{equation}
M=M_{j}=2^{j}\text{ if }I_{j}\leq I<I_{j+1},~j=1,...N  \label{adapt_rule}
\end{equation}%
The region boundaries $\left\{ I_{j}\right\} $ are set to the required
values of turbulence-induced fading required to achieve the target $P_{o}$.
Hence, according to (\ref{BER_M=2}) and (\ref{BER_M>2}), they are obtained by%
\begin{equation}
I_{1}=\sqrt{\frac{1}{2\bar{\gamma}}}Q^{-1}\left( P_{o}\right) ,  \label{thr1}
\end{equation}%
\begin{equation}
I_{j}=\frac{1}{\sin \frac{\pi }{M_{j}}}\sqrt{\frac{1}{2\bar{\gamma}}}%
Q^{-1}\left( \frac{\log _{2}M_{j}}{2}P_{o}\right) ,~j=2,...N  \label{thrN}
\end{equation}%
and%
\begin{equation}
I_{N+1}=R,  \label{infinity}
\end{equation}%
where $R\rightarrow \infty $ and $Q^{-1}\left( \cdot \right) $ denotes the
inverse $Q$-function, which is a standard built-in function in most of the
well-known mathematical software packages. It should be noted that in the
case of $I<I_{1}$, the proposed transmission technique stops transmission.

\subsection{Performance Evaluation}

\subsubsection{Achievable Spectral Efficiency}

The achievable spectral efficiency is defined as the information rate
transmitted in a given bandwidth, and for the communication system under
consideration is given by\footnote{%
Note that since the subcarrier PSK modulation requires twice the bandwidth
than PAM signalling, the average number of bits transmitted in a symbol's interval will be divided by two.}%
\begin{equation}
S=\frac{C}{W}=\frac{\bar{n}}{2}  \label{Spectral1}
\end{equation}%
with $C$ representing the capacity measured in bit/s and $\bar{n}$ the
average number of transmitted bits.

The average number of transmitted bits in the adaptive S-PSK scheme is
obtained by%
\begin{equation}
\bar{n}=\sum_{j=1}^{N+1}a_{j}\log _{2}M_{j}  \label{number_transmitted_bits}
\end{equation}%
where%
\begin{equation}
a_{j}=\Pr \left\{ I_{j}\leq I<I_{j+1}\right\}
=\int_{I_{j}}^{I_{j+1}}f_{I}\left( I\right) dI.  \label{probability}
\end{equation}%
Taking into consideration that the CDF of the LN\ fading distribution is
given by%
\begin{equation}
F_{I}\left( I_{th}\right) =\int_{0}^{I_{th}}f_{I}\left( I\right)
dI=1-Q\left( \frac{\ln \left( I_{th}\right) +2\sigma _{x}^{2}}{2\sigma _{x}}%
\right) ,
\end{equation}%
eq. (\ref{probability}) can be equivalently written as%
\begin{equation}
a_{j}=F_{I}\left( I_{j+1}\right) -F_{I}\left( I_{j}\right) =Q\left(
x_{j}\right) -Q\left( x_{j+1}\right)
\end{equation}%
with $x_{j}=\frac{\ln \left( I_{j}\right) +2\sigma _{x}^{2}}{2\sigma _{x}}$.
Hence, the achievable spectral efficiency can be written as%
\begin{equation}
S=\frac{\sum_{j=1}^{N+1}a_{j}\log _{2}M_{j}}{2}=\frac{\sum_{j=1}^{N}Q\left(
x_{j}\right) -\left( N+1\right) Q\left( x_{N+1}\right) }{2},
\label{spectral2}
\end{equation}%
which is simplified to
\begin{equation}
S=\frac{\sum_{j=1}^{N}Q\left( x_{j}\right) }{2}  \label{spectral3}
\end{equation}%
since $Q\left( x_{N+1}\right) \rightarrow 0$, according to (\ref{infinity}).

\subsubsection{Average Bit Error Rate}

The average BER of the proposed variable rate FSO system can be calculated
as the ratio of the average number of bits in error over the total number of
transmitted bits \cite{J:Aluini_A}. The average number of bits in error can
be obtained by
\begin{equation}
\bar{n}_{err}=\sum_{j=1}^{N+1}\left\langle P_{b}\right\rangle _{j}\log
_{2}M_{j}  \label{err_bits}
\end{equation}%
where
\begin{equation}
\left\langle P_{b}\right\rangle _{j}=\int_{I_{j}}^{I_{j+1}}P_{b}\left(
M_{j},I\right) f_{I}\left( I\right) dI.  \label{BER_mod}
\end{equation}%
Hence the average BER is given by%
\begin{equation}
\bar{P}_{b}=\frac{\bar{n}_{err}}{\bar{n}}.  \label{average_BER}
\end{equation}

\subsection{Application to MIMO FSO systems}

Consider a Multiple Input Multiple Output (MIMO) FSO system where the
information signal is transmitted via $F$ apertures and received by $L$
apertures. For the MIMO system under consideration, it is assumed that the
information bits are modulated using S-PSK and transmitted through the $F$
apertures using repetition coding \cite{J:Saf}. Thus, the received block of
symbols at the $l$th receive aperture is given by%
\begin{equation}
r_{l}\left[ k\right] =\frac{\eta \mu P\sqrt{E_{g}}s\left[ k\right] }{FL}%
\sum_{f=1}^{F}I^{(fl)}+\frac{1}{L}n\left[ k\right] ,~k=1,..K
\label{receive_branch}
\end{equation}%
where $I^{\left( fl\right) }$ denotes the fading coefficient that models the
atmospheric turbulence through the optical channel between the $f$th
transmit and the $l$th receive aperture and $n\left[ k\right] $ represents
AWGN with $\mathrm{E}\left\{ n\left[ k\right] n^{\ast }\left[ k\right]
\right\} =2\sigma _{n}^{2}$. After equal gain combining (EGC) the optical
signals from the $L$ receive apertures, the output of the receiver will be
obtained as%
\begin{equation}
r\left[ k\right] =\sum_{l=1}^{L}r_{l}\left[ k\right] =\frac{\eta \mu P\sqrt{%
E_{g}}s\left[ k\right] }{FL}\sum_{f=1}^{F}\sum_{l=1}^{L}I^{(fl)}+n\left[ k%
\right] .  \label{Eq:EGC}
\end{equation}%
Note that the factor $F$ that appears in (\ref{receive_branch}) and (\ref%
{Eq:EGC}), is included in order to ensure that the total transmit power is
the same with that of a system with no transmit diversity, while the factor $%
L$ ensures that the sum of the $L$ receive aperture areas is the same with
the aperture area of a system with no receive diversity. Moreover, the
statistics of the fading coefficients of the underlying FSO links are
considered to be statistically independent; an assumption which is realistic
by placing the transmitter and the receiver apertures just a few centimeters
apart \cite{J:Lee}.

The variable rate subcarrier PSK transmission scheme can be directly applied
to the MIMO configuration with the decision on the modulation order to be
based on%
\begin{equation}
I_{T}=\frac{\sum_{f=1}^{F}\sum_{l=1}^{L}I^{(fl)}}{FL}.  \label{metr_EGC}
\end{equation}%
Hence, after determining the region boundaries for the target $P_{o}$
requirement, using Eqs. (\ref{thr1})-(\ref{infinity}), the optimum
modulation order will be selected from the $N$ available ones depending on
the value of $I_{T}$, i.e.%
\begin{equation}
M=M_{j}\text{ if }I_{j}\leq I_{T}<I_{j+1},~j=1,...N
\end{equation}

Similarly to the Single Input Single Output (SISO)\ configuration, the
achievable spectral efficiency of the adaptive MIMO FSO system will be
obtained by
\begin{equation}
S=\frac{\sum_{j=1}^{N+1}b_{j}\log _{2}M_{j}}{2}  \label{spectr_div}
\end{equation}%
where
\begin{equation}
b_{j}=\Pr \left\{ I_{j}\leq I_{T}<I_{j+1}\right\}
=\int_{I_{j}}^{I_{j+1}}f_{I_{T}}\left( I\right) dI  \label{diversity_b}
\end{equation}%
and $f_{I_{T}}\left( I\right) $ is the the PDF of $I_{T}$, which can be
efficiently approximated by the lognormal distribution \cite{J:Navid}, \cite%
{J:Lee}, i.e.
\begin{equation}
I_{T}=\exp \left( \xi \right)
\end{equation}%
with $f_{\xi }\left( \xi \right) =N\left( m_{\xi },\sigma _{\xi }^{2}\right)
$, $\sigma _{\xi }^{2}=\ln \left( 1+\frac{e^{4\sigma _{x}^{2}}-1}{FL}\right)
$ and $m_{\xi }=-\frac{1}{2}\sigma _{\xi }^{2}$. Hence, Eq. (\ref%
{diversity_b}) can be equivalently written as
\begin{equation}
b_{j}=Q\left( y_{j}\right) -Q\left( y_{j+1}\right)
\end{equation}%
with $y_{j}=\frac{\ln \left( I_{j}\right) +\frac{1}{2}\sigma _{\xi }^{2}}{%
\sigma _{\xi }}$ and (\ref{spectr_div}) is reduced to
\begin{equation}
S=\frac{\sum_{j=1}^{N}Q\left( y_{j}\right) }{2}.
\end{equation}

Furthermore, the average BER of the adaptive MIMO FSO system will be
obtained by
\begin{equation}
\bar{P}_{b}=\frac{\sum_{j=1}^{N+1}\left\langle P_{b}\right\rangle _{j}\log
_{2}M_{j}}{\sum_{j=1}^{N+1}b_{j}\log _{2}M_{j}}  \label{ber_div}
\end{equation}%
where
\begin{equation}
\left\langle P_{b}\right\rangle _{j}=\int_{I_{j}}^{I_{j+1}}P_{b}\left(
M_{j},I\right) f_{I_{T}}\left( I\right) dI.  \label{diversity_BER}
\end{equation}

\section{Results \& Discussion\label{Results}}

In this section, we present numerical results for the performance of the
adaptive S-PSK scheme in various turbulence conditions and for different
target BERs. We further apply this transmission policy to different MIMO
deployments.

\begin{figure}[tbp]
\centering\includegraphics[keepaspectratio,width=6in]{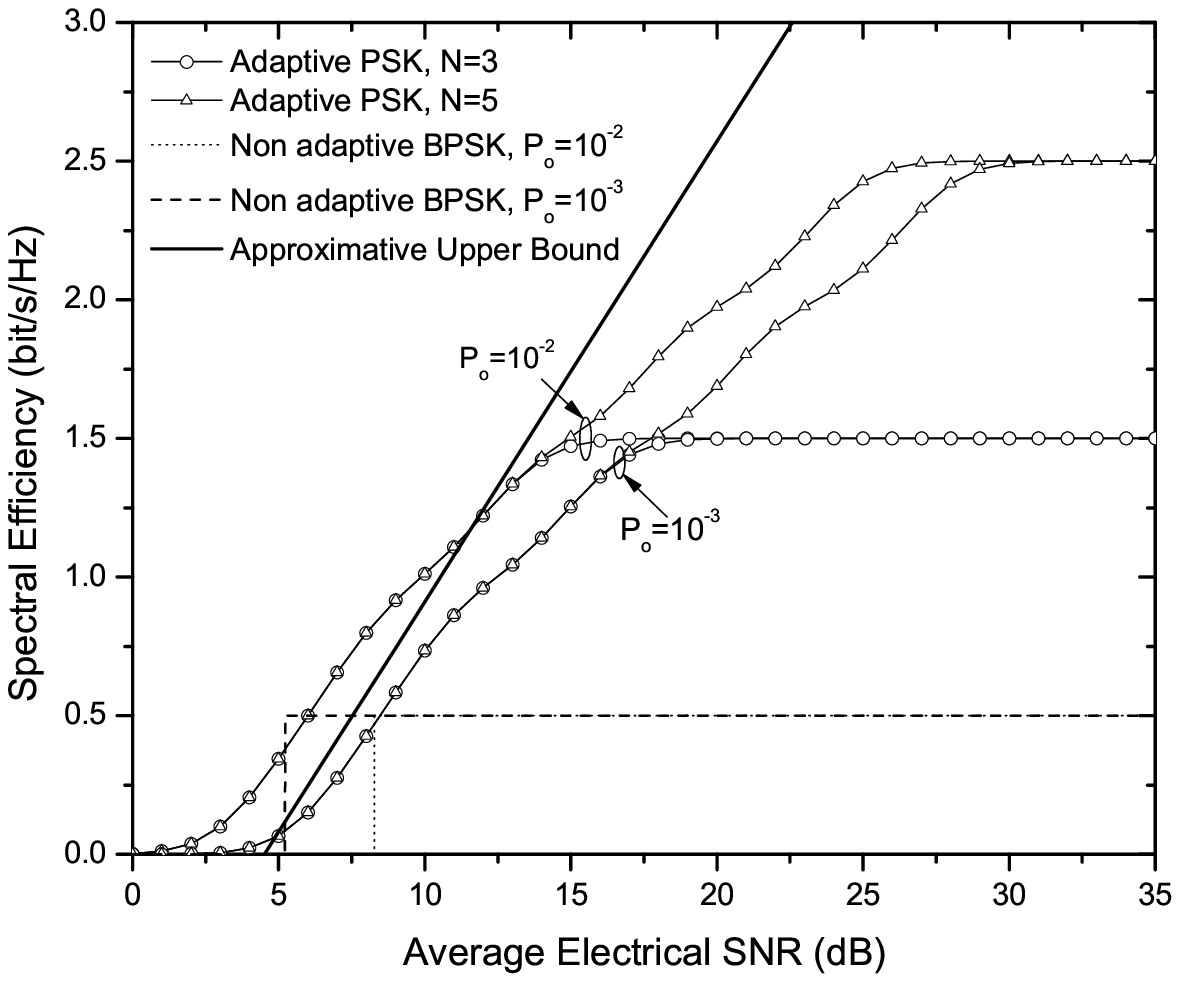}
\caption{Spectral efficiency of the adaptive subcarrier PSK scheme when $%
\protect\sigma _{x}=0.1$.}
\label{Fig:figure1}
\end{figure}

\begin{figure}[tbp]
\centering\includegraphics[keepaspectratio,width=6in]{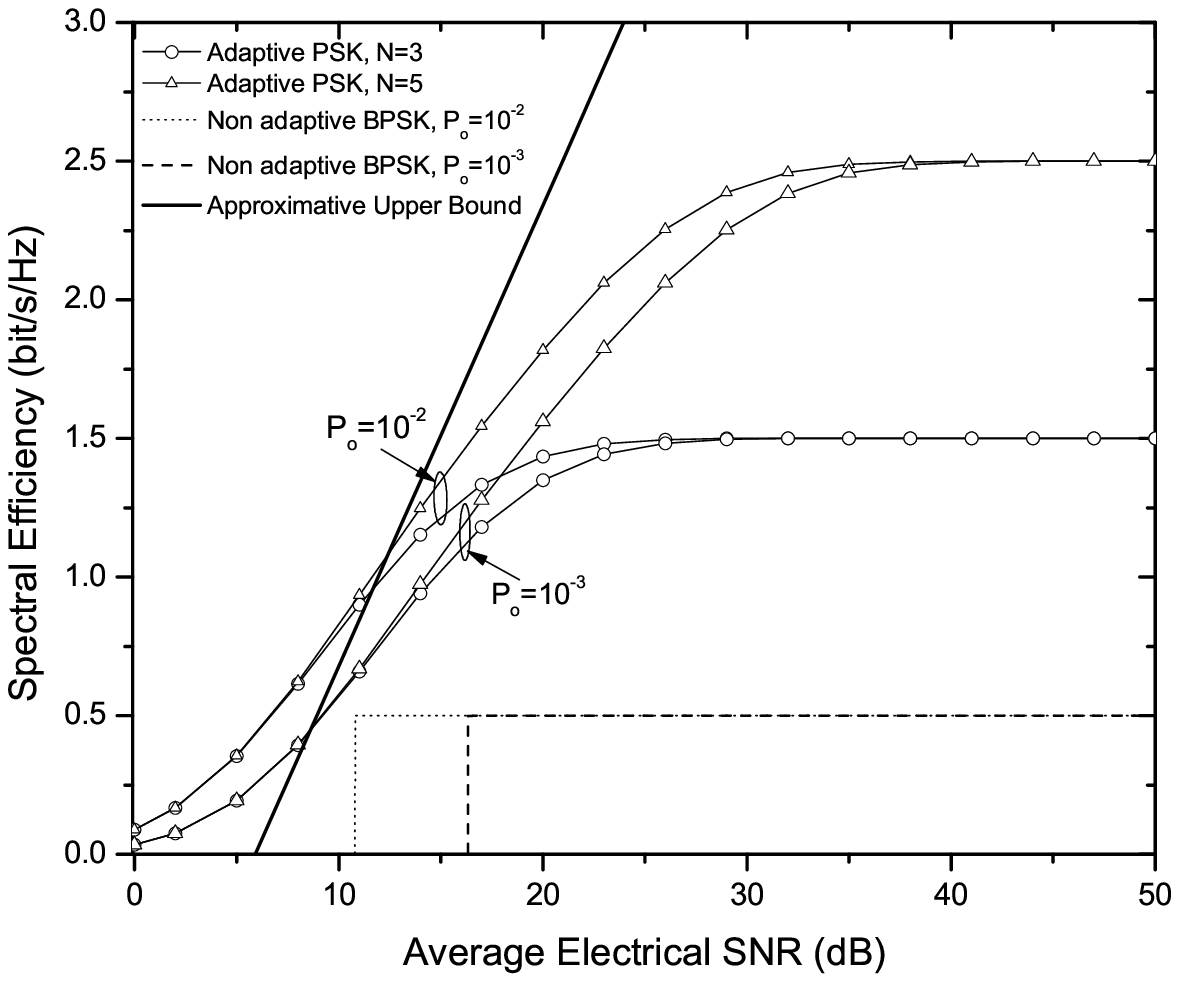}
\caption{Spectral efficiency of the adaptive subcarrier PSK scheme when $%
\protect\sigma _{x}=0.3$.}
\label{Fig:figure2}
\end{figure}

\begin{figure}[tbp]
\centering\includegraphics[keepaspectratio,width=6in]{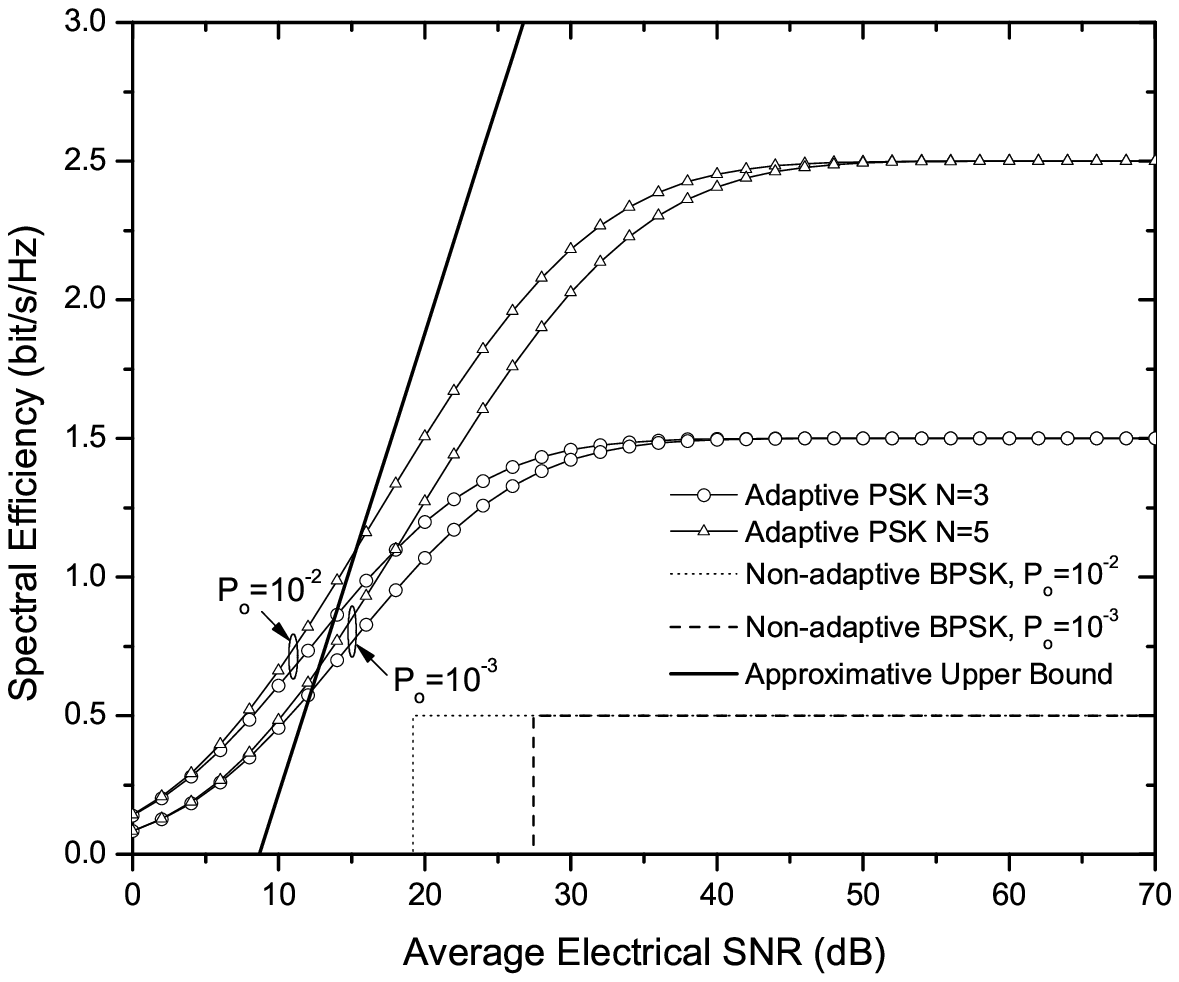}
\caption{Spectral efficiency of the adaptive subcarrier PSK scheme when $%
\protect\sigma _{x}=0.5$.}
\label{Fig:figure3}
\end{figure}

Figs. \ref{Fig:figure1}-\ref{Fig:figure3} depict the spectral efficiency of
the adaptive subcarrier PSK transmission scheme at different degrees of
turbulence strength, i.e. $\sigma _{x}=0.1$, $\sigma _{x}=0.3$ and $\sigma
_{x}=0.5$, and when a SISO\ FSO system is considered. Specifically,
numerical results obtained by (\ref{spectral3}) for two different target BER
requirements, $P_{o}=10^{-2}$ and $P_{o}=10^{-3}$, are plotted as a function
of the average electrical SNR, along with the upper bound provided by (\ref%
{uncond_cap2}) and the spectral efficiency of the non-adaptive subcarrier
BPSK ($M=2$). The latter is found by determining the value of the average
electrical SNR for which the BER performance of the non-adaptive BPSK, as
given by (\ref{uncond_BER}), equals $P_{o}$. It is obvious from the figures
that the spectral efficiency of the adaptive transmission scheme increases
and comes closer the upper bound by increasing the target $P_{o}$. Moreover,
when compared to the non-adaptive BPSK, it is observed that adaptive PSK
offers large spectral efficiency gains (14dB when $P_{o}=10^{-3}$) at strong
turbulence conditions ($\sigma _{x}=0.5$); however, these gains are reduced
as $\sigma _{x}$ reduces. For low turbulence ($\sigma _{x}=0.1$), it is
observed that non-adaptive BPSK reached its maximum spectral efficiency ($%
S=0.5$) at lower values of SNR than the proposed adaptive scheme, indicating
that in these conditions it is more effective to modify the modulation order
based on the value of the average SNR rather than the instant value of
fading intensity. Hence, the proposed adaptive PSK technique, which is based
on the estimation of the instantaneous value of the channel's fading
intensity, can be considered particularly effective only in the moderate to
strong turbulence regime.

\begin{figure}[tbp]
\centering\includegraphics[keepaspectratio,width=6in]{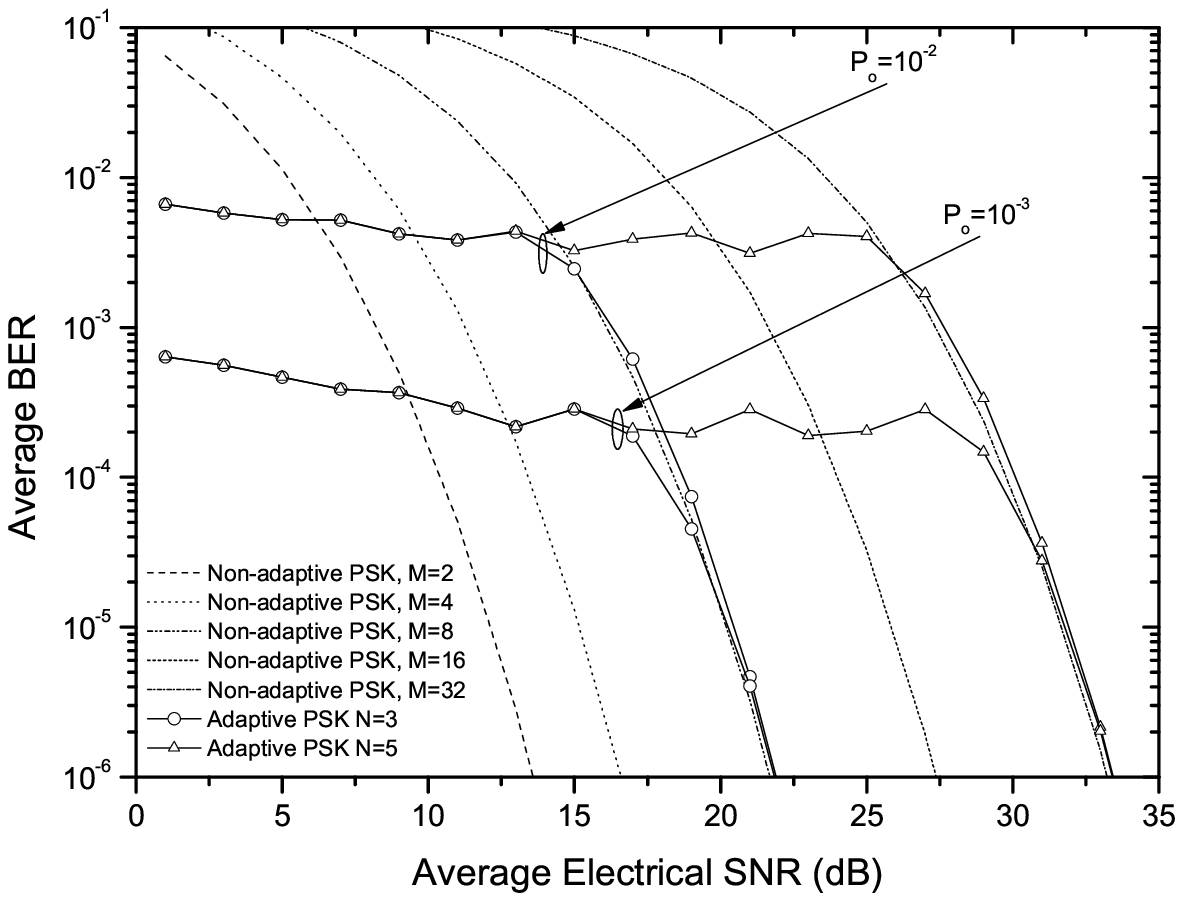}
\caption{Average BER of the adaptive subcarrier PSK scheme when $\protect%
\sigma _{x}=0.1$.}
\label{Fig:figure4}
\end{figure}

\begin{figure}[tbp]
\centering\includegraphics[keepaspectratio,width=6in]{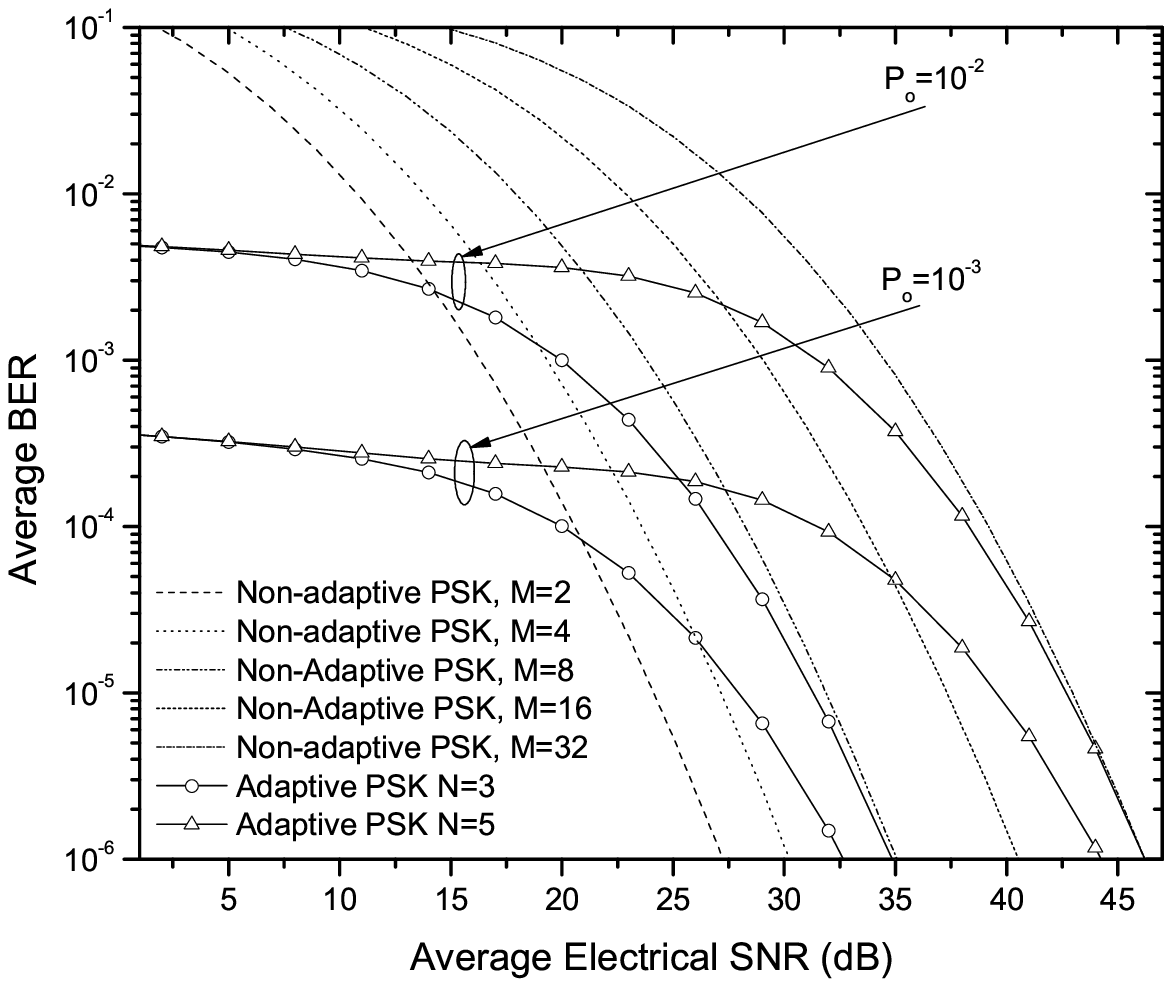}
\caption{Average BER of the adaptive subcarrier PSK scheme when $\protect%
\sigma _{x}=0.3$.}
\label{Fig:figure5}
\end{figure}

\begin{figure}[tbp]
\centering\includegraphics[keepaspectratio,width=6in]{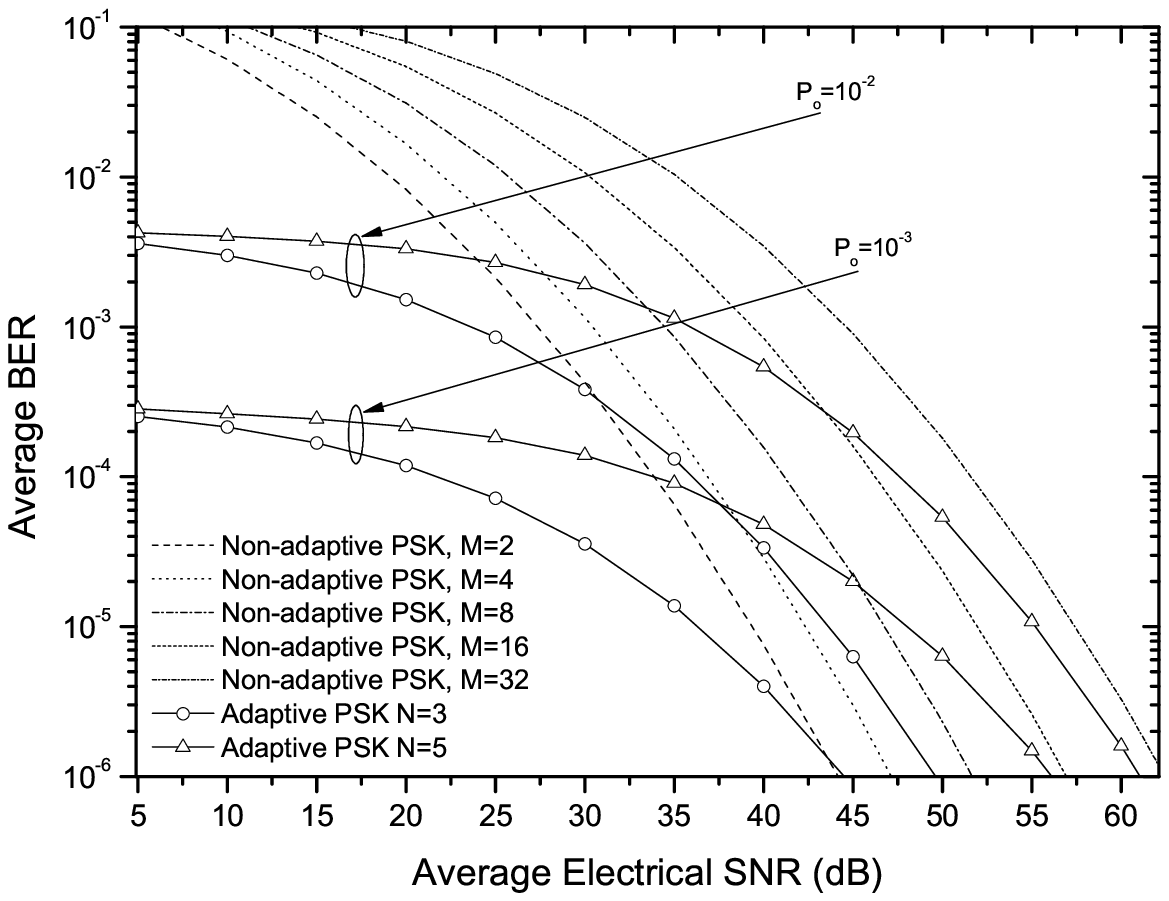}
\caption{Average BER of the adaptive subcarrier PSK scheme when $\protect%
\sigma _{x}=0.5$.}
\label{Fig:figure6}
\end{figure}

Figs. \ref{Fig:figure4}-\ref{Fig:figure6} illustrate the average BER
performance of the adaptive transmission technique for the same target BER
requirements and turbulence conditions. It is clearly depicted that the
average BER of the adaptive system is lower than the target $P_{o}$ in all
cases examined, satisfying the basic design requirement of (\ref%
{optimization problem}). Moreover, it can be easily observed that the
performance of the adaptive system approaches the performance of the
non-adaptive system with the largest modulation order, at high values of
average SNR; this was expected, since in this SNR regime the adaptive scheme
chooses to transmit with the largest available modulation order.

\begin{figure}[tbp]
\centering\includegraphics[keepaspectratio,width=6in]{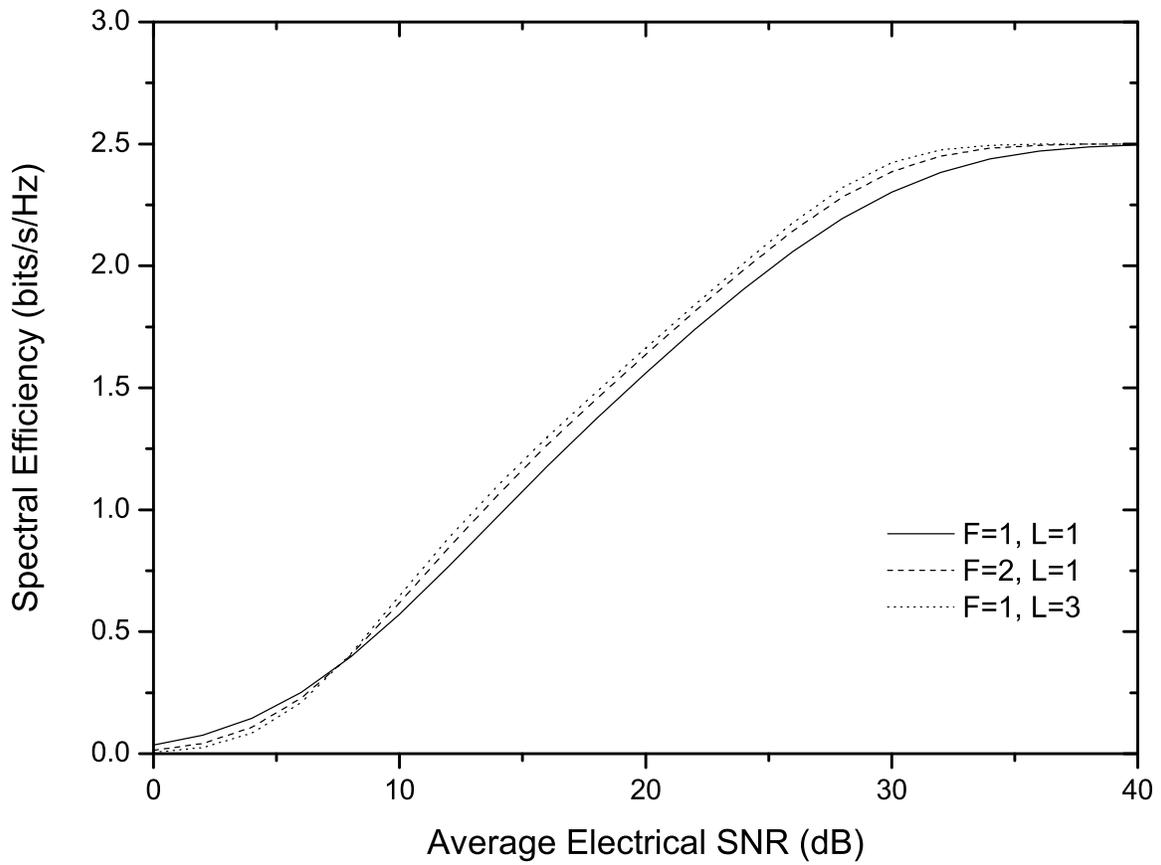}
\caption{Spectral efficiency of the adaptive subcarrier PSK scheme for
various MIMO configurations, when $N=5$, $P_{o}=10^{-3}$ and $\protect\sigma %
_{x}=0.3$.}
\label{Fig:figure7}
\end{figure}

Finally, Fig. \ref{Fig:figure7} depicts numerical results for the spectral
efficiency of various MIMO FSO systems, when the adaptive transmission
technique with target $P_{o}=10^{-3}$ and $N=3$ available modulation orders
is applied and moderate turbulence conditions are considered. As it is
clearly illustrated in the figure, the increase of the number of transmit
and/or receive apertures improves the performance of the adaptive
transmission scheme, increasing the achievable spectral efficiency. However,
this does not happen at low values of average SNR (less than 8dB), which may
seem surprising at first but can be explained by the following argument. At
the low average SNR regime, the region boundaries $\left\{ I_{j}\right\} $
correspond to values higher than the unity. Hence, as the number of transmit
and/or receive apertures increases, the variance $\sigma _{\xi }^{2}$
decreases, resulting in higher values for the parameters $\left\{
y_{j}\right\} $ and decreased spectral efficiency. As the average SNR
increases, most of the region boundaries $\left\{ I_{j}\right\} $ take
values lower than unity and, as a concequence, the increase in the number of
apertures results in lower values for the parameters $\left\{ y_{j}\right\} $
and higher spectral efficiency.

\section{Conclusions\label{Conclusions}}

We have presented a novel adaptive transmission technique for FSO systems
operating in atmospheric turbulence and employing S-PSK intensity
modulation. The described technique was implemented through the modification
of the modulation order of S-PSK according to the instantaneous fading state
and a pre-defined BER requirement. Novel expressions for the spectral
efficiency and average BER of the adaptive FSO system were derived and
investigations over various turbulence conditions and target BER
requirements were performed. Numerical results indicated that adaptive
transmission offers significant spectral efficiency gains, compared to the
non-adaptive modulation, at the moderate-to-strong turbulence regime (14dB
at $S=0.5$, when $P_{o}=10^{-3}$ and $\sigma _{x}=0.5$); however, it was
observed that in low turbulence, it is more efficient to perform adaptation
based on the average electrical SNR, instead of the instantaneous fading
state. Furthermore, the proposed technique was applied at MIMO FSO systems
and additional improvement in the achieved spectral efficiency was observed
at the high SNR regime, as the number of the transmit and/or receive
apertures increased.

\section*{Appendix}

This appendix provides a closed-form expression for (\ref{uncond_cap}).
Using the pdf of turbulence induced fading, (\ref{uncond_cap}) can be
written as
\begin{equation}
C_{up}\approx \frac{W}{2}\int_{0}^{\infty }\log _{2}\left( \frac{\bar{\gamma}%
I^{2}}{e}\right) \frac{1}{2I\sqrt{2\pi \sigma _{x}^{2}}}\exp \left( -\frac{%
\left( \ln I+2\sigma _{x}^{2}\right) ^{2}}{8\sigma _{x}^{2}}\right) dI.
\label{uncond_cap3}
\end{equation}%
To simplify (\ref{uncond_cap3}), we substitute $\ln I$ by $y$ and hence%
\begin{equation}
C_{up}\approx K_{1}+K_{2}  \label{uncond_cap4}
\end{equation}%
where
\begin{equation}
K_{1}=\frac{W}{4\sqrt{2\pi \sigma _{x}^{2}}}\log _{2}\left( \frac{\bar{\gamma%
}}{e}\right) \int_{-\infty }^{\infty }\exp \left( -\frac{\left( y+2\sigma
_{x}^{2}\right) ^{2}}{8\sigma _{x}^{2}}\right) dy  \label{K1}
\end{equation}%
and%
\begin{equation}
K_{2}=\frac{W}{2\ln 2\sqrt{2\pi \sigma _{x}^{2}}}\int_{-\infty }^{\infty
}y\exp \left( -\frac{\left( y+2\sigma _{x}^{2}\right) ^{2}}{8\sigma _{x}^{2}}%
\right) dy.  \label{K2}
\end{equation}%
Using \cite[Eq. (3.321/3)]{B:Gra_Ryz_Book}, (\ref{K1}) is reduced to
\begin{equation}
K_{1}=\frac{W}{2}\log _{2}\left( \frac{\bar{\gamma}}{e}\right)
\label{K1_red}
\end{equation}%
while, using \cite[Eq. (3.461/2)]{B:Gra_Ryz_Book}, (\ref{K2}) is reduced to%
\begin{equation}
K_{2}=-\frac{2W\sigma _{x}^{2}}{\ln 2}.  \label{K2_red}
\end{equation}%
Hence, by substituting in (\ref{uncond_cap4}), Eq. (\ref{uncond_cap2}) is
obtained.

\enlargethispage{-0.2in}

\newpage
\newpage
\newpage
\newpage
\newpage
\newpage
\newpage

\end{document}